\documentclass[aps,prl,twocolumn,superscriptaddress,nopacs]{revtex4-1}

\usepackage{graphicx}
\usepackage{amssymb}
\usepackage{amsmath}
\usepackage{epstopdf}
\usepackage{mciteplus}
\usepackage{setspace}

\begin{document}

\title{Formation of shear-bands in drying colloidal dispersions}

\author{Pree-Cha Kiatkirakajorn}
\author{Lucas Goehring}
\email[]{lucas.goehring@ds.mpg.de}
\affiliation{Max Planck Institute for Dynamics and Self-Organization (MPIDS), 37077 G\"ottingen, Germany}

\date{\today}

\begin{abstract}
In directionally-dried colloidal dispersions regular bands can appear behind the drying front, inclined at $\pm45^\circ$ to the drying line.  Although these features have been noted to share visual similarities to shear bands in metal, no physical mechanism for their formation has ever been suggested, until very recently.  Here, through microscopy of silica and polystyrene dispersions, dried in Hele-Shaw cells, we demonstrate that the bands are indeed associated with local shear strains.  We further show how the bands form, that they scale with the thickness of the drying layer, and that they are eliminated by the addition of salt to the drying dispersions.  Finally, we reveal the origins of these bands in the compressive forces associated with drying, and show how they affect the optical properties (birefringence) of colloidal films and coatings.
\end{abstract} 


\maketitle

Colloidal dispersions are the basis of many paints, inks and coatings, are used in ceramics and composites, and are a foundation of the photonics industry.   When dried, these dispersions can show a surprising variety of patterning mechanisms.  They may buckle~\cite{Tsapis2005}, or crack into spirals~\cite{Lazarus2011}, waves~\cite{Goehring2011}, or parallel lines~\cite{Allain1995,Dufresne2006}, for example, while their dried shape can vary from a coffee-ring~\cite{Deegan1997} to textured surfaces controlled by evaporation-lithography \cite{Harris2007,Parneix2010}.  This abundance of patterns, recently reviewed in \cite{Routh2013,Thiele2014}, suggests diverse means for the directed self-assembly of micro-structured materials, if the underlying dynamics can be understood and controlled.  

\begin{figure}
\includegraphics[width=3.375in]{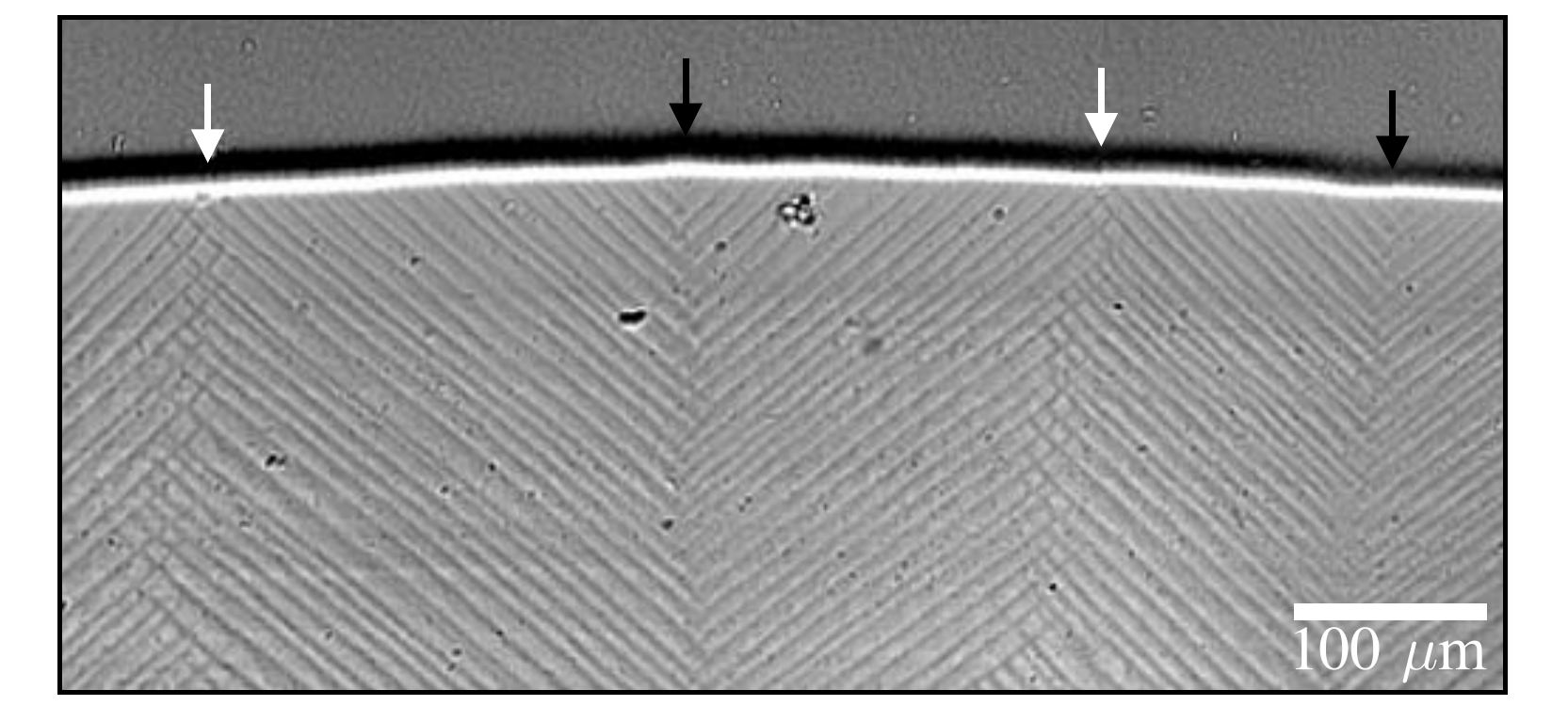}
\caption{Drying colloidal polystyrene in a free-standing film.  The drying front is a bright horizontal line, with liquid film above it, and solid film below.  The shear bands in the solid are angled away from the drying front.  Domains of left-leaning and right-leaning bands can meet in two ways: junctions that open towards the drying front (black arrows) are sources of new bands, while those that open away from the front (white arrows) are where growing bands end. \label{bandpic}}
\end{figure}

This letter focusses on establishing and manipulating the driving forces behind a banded (or striped) structure that is commonly found in drying dispersions, and shown in Fig. \ref{bandpic}.  Such bands were first noticed in sol-gels by Hull and Caddock \cite{Hull1999}, who studied desiccation as an analog for thermal contraction in geophysical settings.  They found that regular bands always preceded fracture, and that the bands were oriented at $\pm45^\circ$ to the cracks.  Similar bands can be seen in the figures of many studies of drying colloids (\textit{e.g.} \cite{Dufresne2006,Goehring2010,Smith2011,Boulogne2014}), but passed without further discussion until the work of Berteloot \textit{et al.}, who described their appearance on the surface of drying colloidal droplets \cite{Berteloot2012}.  They suggested that the bands might be either shear bands, based on a visual similarity to these features in metals \cite{Schroers2004}, or the surface buckling of a colloidal skin.  

Recently, Yang \textit{et al.} have shown that these bands are indeed linked to the shear-response of a drying film \cite{Yang2015}.  They studied free-standing films of colloidal polystyrene, and effectively controlled the yield stress and strain-rate (drying rate) of the films.  In particular, they showed how the spacing and relative \textit{widths} of the bands -- \textit{i.e} the fraction of sheared material -- agrees with the lever rule of shear localisation \cite{Yang2015}.  This rule, derived from the physics of two-phase coexistence, describes the instability by which shear localises into finite bands in a variety of complex fluids \cite{Ovarlez2009}.

Here we give a complementary view of these shear bands, identifying the forces behind their formation.  We show by direct means that they are the result of shear deformations, and further demonstrate how their pattern scales, how it can be suppressed, and discuss its origins in the compressive forces that accompany the directional drying of colloidal dispersions.  In contrast to \cite{Berteloot2012,Yang2015}, we study slow drying in Hele-Shaw cells, to eliminate any possible influence of a free surface or skin.

To prepare banded films, we used a variety of charge-stabilized dispersions.  Polystyrene dispersions with particle diameters $d$ of 98, 100, 105, 115, 144, 198 and 283 nm were synthesized as described elsewhere \cite{Yow2010}.  For colloidal silica, Ludox HS (Sigma-Aldrich, $d = 16$ nm) was dialyzed against aqueous solutions of 5 mM NaCl at pH 9.5 (by addition of NaOH), while Levasil-30 (Azko-Nobel, $d = 92$ nm \cite{Boulogne2014}) was used as received.  NaCl solutions were then mixed with the above materials to form dispersions of $\sim 5\%$ solids by volume, and salt concentrations of up to 60 mM.  Deionized water (Millipore) was used for all procedures.  

\begin{figure}
\includegraphics[width=3.375in]{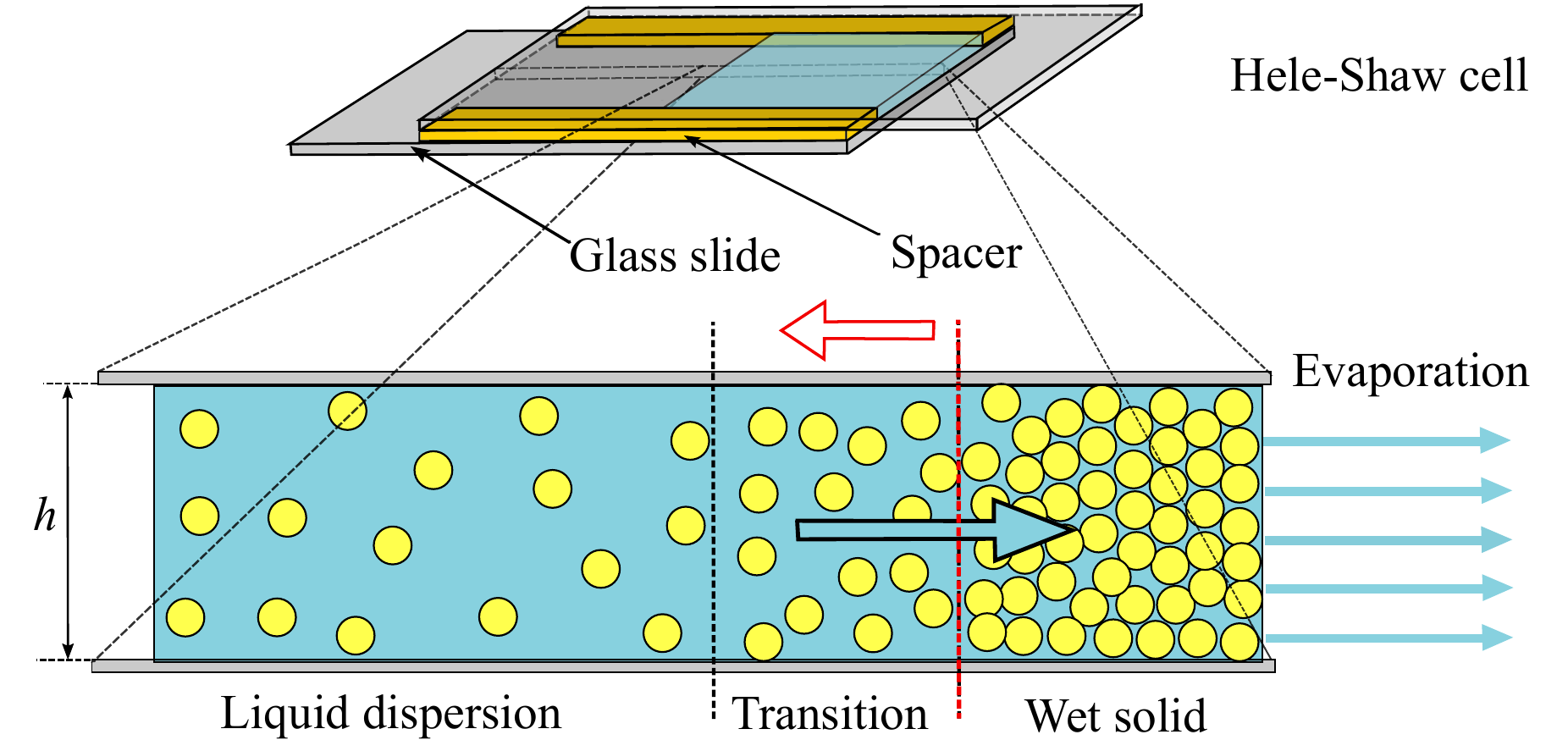}
\caption{Drying cells were made in a Hele-Shaw geometry.  A dispersion was drawn into a cell and then dried naturally from one edge. As drying proceeded, a solid region slowly grew into the still-liquid dispersion.  On close inspection the solidification front is a narrow transition region with strong gradients in particle concentration, over which the dispersion's properties rapidly change.  \label{cell}}
\end{figure}

Drying experiments were performed in Hele-Shaw cells built from two 2.5$\times$7.5 cm glass microscope slides, as in Fig. \ref{cell}.  Interposed between the slides, along their long edges, were plastic spacers of either 150, 230, 250, 300, 400 or 500 $\mu$m thickness.  The cells were held together by clips and their actual thicknesses $h$ were measured by microscopy after assembly.  Thinner cells, with $h$ between 36--72 $\mu$m, were made with heat-curable polymer sheets or double-sided tape as spacers.  In each experiment a dispersion was pipetted into the cavity of a cell.  The sample then dried by evaporation from one of the open short edges of the cell, at room temperature.  During drying the cell was slightly inclined by a few degrees, so that any bubbles rose away from the drying edge by buoyancy.  

The drying of these dispersions proceeds by the advance of a solidification front into the colloidal liquid (see {\it e.g.} \cite{Allain1995,Dufresne2003,Goehring2010,Routh2013}), as sketched in Fig. \ref{cell}.  The solid film remains porous and water flows through it to balance evaporation from its open edge.  This flow continuously brings more colloidal particles with it, which add to the growing solid deposit.   The viscous drag of the water flowing past these particles is balanced by a gradient in their osmotic pressure, and hence a concentration gradient forms in the liquid leading up to the solidification front.   Along this gradient the character of the dispersion can change rapidly, and dramatically.   For example, in charge-stabilized dispersions, as those here, the particles may first be concentrated into a soft repulsive solid, where each particle is caged by long-range electrostatic interactions with its neighbors \cite{Goehring2010}.  The drag from the flow through this soft solid is equivalent to a uniaxial compression and can cause effects like structural anisotropy and birefringence \cite{Boulogne2014}. Eventually the particles are pushed close enough together for short-range van der Waals forces to overcome their electrostatic repulsion.  At this point the particles will jump into intimate contact, forming into a cohesive, rigid film \cite{Goehring2010}.  

We found shear bands in our dried colloidal films, in both Hele-Shaw geometries and in free-standing droplets dried from the same materials.   Generally, as shown in Fig. \ref{bandpic} (free-standing), or in the supplemental movie \cite{SI} (Hele-Shaw), the bands appeared as a dispersion changed from a liquid to a solid.   They arranged into domains of stripes inclined to either the left or the right of the drying front. These domains overlapped by about one band spacing, forming domain boundaries that looked like chevrons \cite{Yang2015} opening either towards the drying front, or away from it.  

The chevrons that opened towards the drying front periodically generated new bands, which appeared alternately to one side of the chevron, and then to the other.  The chevrons that opened away from the drying front were sinks for growing bands.  New bands could also appear between two existing bands, if they were far enough apart.  A new band typically began at a point and generated two `tips' that propagated simultaneously towards and away from the drying front, over several seconds.   Banding always occurred well before any cracking.

\begin{figure}
\includegraphics[width=85mm]{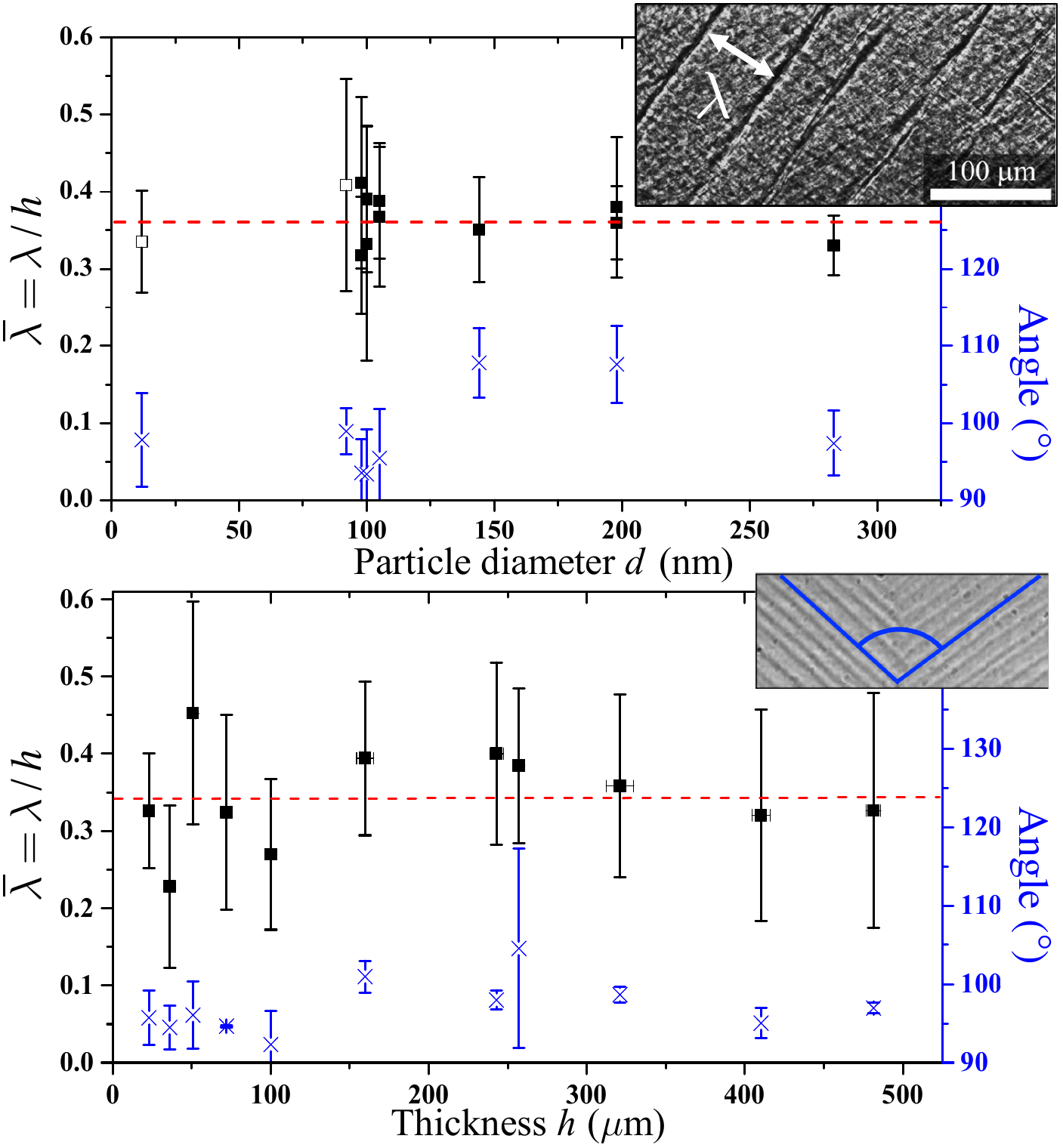}
\caption{The relative band spacing $\bar\lambda$ (squares, left axis) and band intersection angle (blue crosses, right axis) were measured in Hele-Shaw cells.   Bars show the standard deviation of each measurement, while inset panels demonstrate how the measurements are made. (a) shows results in $150$ $\mu$m thick cells, for colloidal latex (filled squares) and silica (open squares) at low ionic strength (1-5 $\mu$M), while (b) shows data for particles with $a\simeq100$ nm, in cells of different thickness.  Neither measurement depends on particle size or thickness; the band spacing is always simply proportional to~$h$.\label{band_spacing}}
\end{figure}

The shear bands approached the liquid-solid boundary at an angle of 45$^\circ$, but were then compressed along the direction of drying over time.  This smooth bending of the bands can be seen near the liquid-solid transition of Fig. \ref{bandpic}, or in the supplemental movie  \cite{SI}.  Later, if a film cracked, the bands were also compressed by the crack opening.  Looking at 21 different \textit{fully-dried} films, in both free-standing and Hele-Shaw geometries, we found that the final opening angle of the chevrons was 99$\pm$7$^\circ$ (mean $\pm$ standard deviation).  There was no clear dependence on particle size or film thickness (Fig. \ref{band_spacing}).

The most obvious feature of the bands is their relatively even spacing, $\lambda$.  We observed bands in a range of fully-dried colloidal materials, and a variety of conditions.  For each of eight different particle sizes we prepared 150 $\mu$m thick Hele-Shaw cells, filled with 180 $\mu$l of dispersion.  Once dry, we imaged these cells by digital microscopy under transmitted light.  We scanned across the width of each cell, collected images in about ten different locations, and measured about 20 band spacings (measured normal to the bands) per image.

Figure \ref{band_spacing}(a) shows that there was no dependence of the band spacing on the particle size, for particles between 16 and 283 nm.  Here $\lambda$ was, on average, 0.36 times the film thickness $h$.  An equal mixture of two particle sizes, 98 and 198 nm, was also consistent ($\lambda/h = 0.29\pm0.13$) with this result. However, although the bands were parallel and had a clearly defined spacing, there could be wide variations in this spacing within a single cell: as in Fig. \ref{bandpic}, patches of wider or narrower bands often appeared together, with no apparent reason for the local variation in size.   As such, we report the standard deviation (rather than the standard error) of our measurements, throughout this letter.   Bands regularly formed with a range of spacings from 0.2 to 0.5 times the film thickness.
 
Dispersions with $d\simeq 100$ nm (each experiment was repeated with 92 nm silica and 105 and 115 nm polystyrene particles) were then dried in cells of different thicknesses. Figure \ref{band_spacing}(b) shows that there was no dependence of the relative band spacing $\bar{\lambda} = \lambda/h$ on the thickness of the drying film.  In other words, the band spacing scales linearly with $h$.  Although there was, again, considerable variation within the results, the average  $\bar{\lambda} = 0.34$ agrees with the results for changing particle size.

\begin{figure}
\includegraphics[width=85mm]{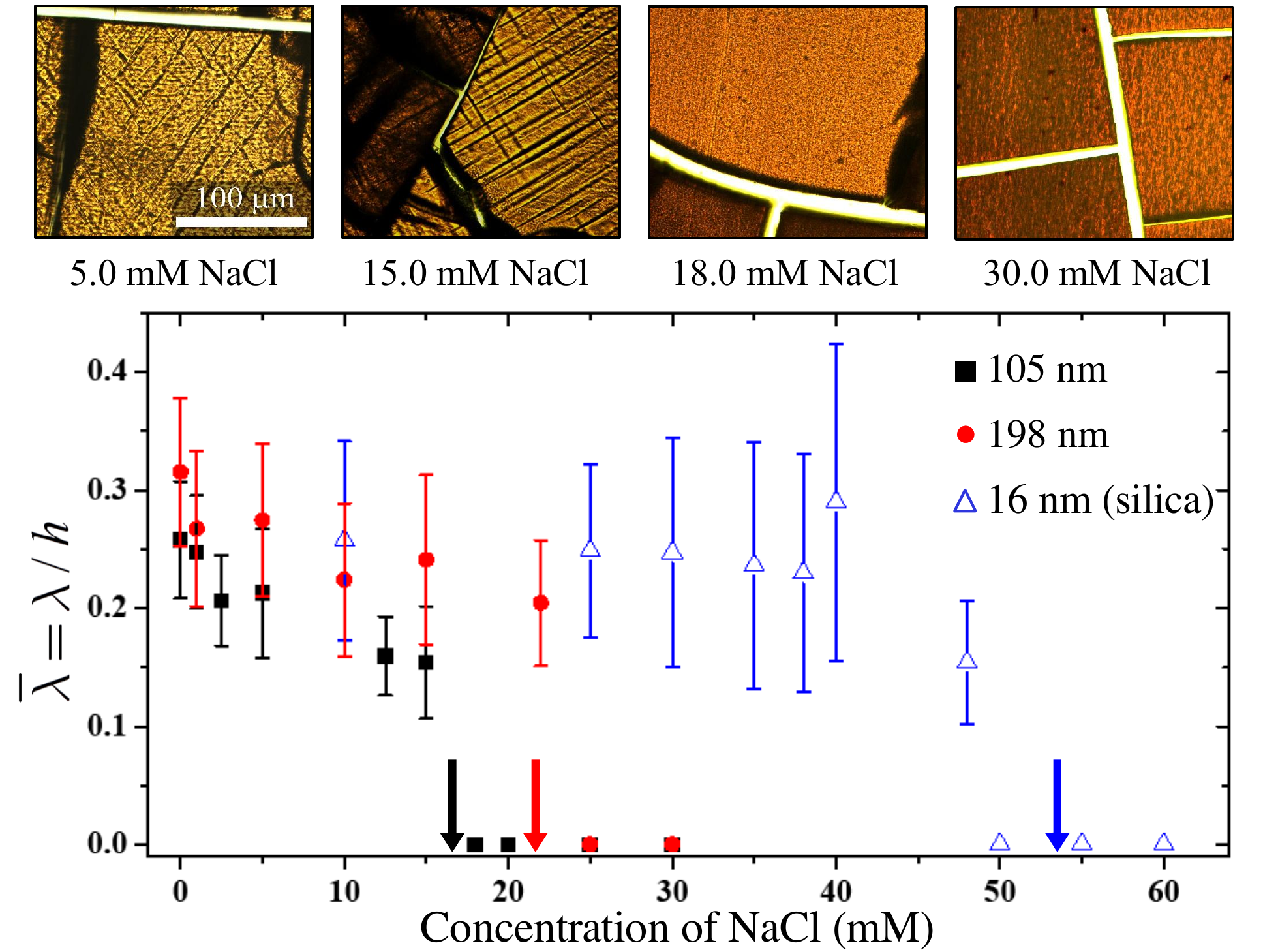}
\caption{Shear bands may be eliminated ($\bar\lambda = 0$ indicates no banding) by the addition of salt, with a critical level depending on particle size and composition.  Arrows mark where the strongest interactions between neighboring particles (Eq. \ref{DLVO}) drop below 5$kT$.  Above, images show how the dry films change with salt concentration, for 105 nm polystyrene. \label{band_spacing3}}
\end{figure}

Next, we studied how the formation of bands was influenced by the salinity of the dispersions.  For charged colloids, dissolved salt screens electrostatic interactions, affecting the cooperative behavior of the particles. We therefore dried 105 and 198 nm colloidal polystyrene and 16 nm colloidal silica in NaCl solutions of to up to 60 mM, in 150 $\mu$m thick cells.  Figure \ref{band_spacing3} shows that for each dispersion there is a cutoff salt concentration above which bands do not form.  This critical concentration depends on particle size and type.  Below the critical concentration there also appears to be a weak negative dependence of the band spacing on salt. Otherwise, however, all these dispersions dried exactly as before, \textit{via} directional solidification, and cracked cleanly.  

These experiments collectively suggest how bands form by an elastic instability, similar to fracture.  They are Mode-II (sliding) cracks, or shear bands, that act to release the shear stresses that accompany directional solidification, just as the opening cracks that often form in drying paints relax in-plane tensions.  

Cracks in a thin elastic film are usually regularly spaced by a small multiple of the film thickness \cite{Allain1995,Bai2000a}.  This is often argued to follow from how they release strain energy.  For example, the height-averaged strains near a straight crack in a thin film, with a rigid boundary, decay away over a distance proportional to $h$ \cite{Xia2000}.    The same mechanics requires that distortions rapidly decay away from a shear band in a film attached to rigid walls, and that the mature band spacing should be proportional to the film thickness, as we have shown it is.  

\begin{figure}
\begin{center}
\includegraphics[width=85mm]{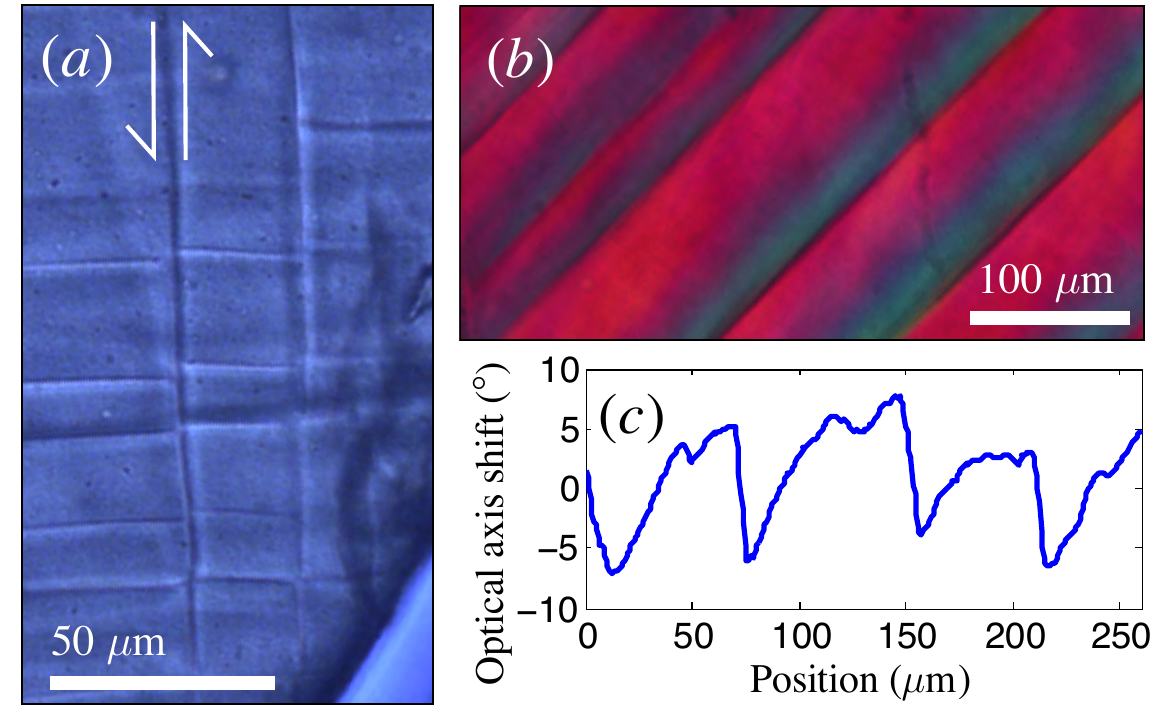}
\caption{Shear deformation.  (a) A later band will displace any earlier bands to either side of it like a fault.  (b) Dry films are birefringent.  Under cross-polarized light a homogenous film should appear purple here.  A clockwise rotation of the optical axis shifts the color towards blue, while the opposite motion shifts it to red.  (c) The optical axis of a film (Levasil) shifts by $\pm$5$^\circ$ around the bands, showing how the entire bulk structure of the film is twisted by band formation.
\label{shear1}}
\end{center}
\end{figure}

During drying, the colloidal dispersion is subjected to a uniaxial compression, caused by the drag of water flowing past the particles to the evaporating edge.  Under these conditions the direction of maximum shear stress is at $\pm$45$^\circ$ to the direction of compression: for principal stresses $\sigma_1$ and $\sigma_2$ the shear stress at an orientation $\theta$ to the compression axis is $((\sigma_2-\sigma_1)/2)\sin(2\theta$).  The von Mises yield criterion (see \textit{e.g.} \cite{Anderson2005}) states that plastic failure is expected when the magnitude of this stress exceeds some critical value, and that failure will occur along the direction of maximum shear stress.  Thus the orientation of the bands, and the position of their formation, show that they relieve the shear stresses generated near the solidification front.   The shear motion around the bands can be seen directly in cases where bands overlap significantly, as in Fig. \ref{shear1}(a), or the supplemental movie \cite{SI}.  Later deformation, including cracking, will distort the inclination of the shear bands, but is not related to their formation.

Directionally dried colloidal films are birefringent, as their entire structure is compressed along the direction of drying \cite{Yamaguchi2013,Boulogne2014}. The optical axis of a material is a direction along which birefringent effects vanish.  It may be thought of as a vector that, on average, lies here normal to the solidification front \cite{Yamaguchi2013}, but which can be locally rotated by any deformations.  The birefringence around the shear bands (Fig. \ref{shear1}(b)) can thus be used to measure how the bands strain the film around themselves.   We positioned our films in a cross-polarizing microscope equipped with a first-order retardation plate, so that the optical axis at a point half-way between the bands was aligned with the polarization of the transmitted light.  We measured the optical axis shift at different points by finding the relative rotation required to minimize transmitted light of 546 nm.  As shown in Fig. \ref{shear1}(c), the formation of a shear band twists the entire structure of the film, rotating the optical axis (\textit{i.e.} the direction of maximum compression of the film) by $\pm$5$^\circ$ to either side of it. The sawtooth-pattern of the birefringence shows how slip is highly localized in the shear band, and how strains decay away smoothly from the slip plane. 

Finally, we discuss the extinction of the shear bands with addition of salt, which shows why band formation occurs only at the liquid-solid transition of directional drying. The DLVO model of colloidal stability combines an attractive van der Waals potential at short distances with a repulsive electrostatic interaction that is screened by ions in solution \cite{Russel1989}.  We use the pair potential
\begin{equation}
U = -\frac{Ad}{24s} + \frac{4dkT}{L_B}\tanh^2(\Phi)e^{-\kappa s}
\label{DLVO}
\end{equation} 
to approximate the free energy of two neighboring particles of surface-separation $s$.  Here, $A$ is the Hamaker constant ($5\cdot10^{-21}$ J for polystyrene \cite{Goehring2010}, and $8\cdot10^{-21}$ J for silica \cite{Russel1989}), $L_B = 0.7$ nm is the Bjerrum length, $kT$ is the Boltzmann energy and $\kappa^{-1}$ is the Debye length. $\Phi$ is a reduced surface potential, calculated as in \cite{Goehring2010} with a surface charge density of 0.53 $\mu$C/cm$^2$ for polystyrene particles (measured in \cite{Goehring2010}), and 2.5 $\mu$C/cm$^2$ for silica (chosen to match the zeta potentials in \cite{Healy2006}).  Adding salt lowers both $\kappa^{-1}$ and $\Phi$.  Experience has shown that colloidal crystals \cite{Russel1989} or glasses \cite{Boulogne2014} will form when a dispersion is concentrated to the point where the pair potential at an average particle separation reaches a few times $kT$.  Above this threshold the repulsive forces between particles can give rise to effective cages, freezing particles into a semi-rigid arrangement. 

For the three cases studied we find that the shear bands disappear at the salt concentrations where the maximum value of $U(s)$ is lowered to $5kT$ (arrows in Fig. \ref{band_spacing3}).  In other words, the bands vanish when the electrostatic interaction is weakened to the point where a soft repulsive solid does not have the opportunity to form during the transition from a colloidal liquid to an aggregated solid film.   Since salt will not significantly affect the properties of the aggregated solid, our analysis demonstrates that it is the compression across the liquid-solid transition itself that is responsible for shear band formation.  

The above model identifies the conditions for an appropriate yield-stress material to form under compressive strains during drying.  A complementary description of how these strains then become localised (\textit{i.e.} the lever rule)  was recently given in \cite{Yang2015}.  We note that particle caging through electrostatic interactions will also become inefficient when $(\kappa d/2) \gg 1$, possibly explaining the absence of bands for the 200 nm particles in that study \cite{Yang2015}. 

Here we have shown how the shear bands that regularly appear in dried colloidal films form and scale. The bands relieve the compressive stresses associated with directional drying, by localized slip at 45$^\circ$ to the axis of compression.  In particular, we have shown how the bands are driven by the compression of a soft intermediate phase that can form around the liquid-solid transition, when particles are caged by their electrostatic interactions.  By engineering the response of this transition region, such as by adding salt to weaken inter-particle forces, the bands can be controlled or eliminated. 

\begin{acknowledgments}
We thank Anupam Sengupta and Carsten Kr\"uger for assistance with the birefringence measurements, and Antoine Fourri\`ere for helpful discussions.  PK thanks the Thai DPST and Government of Thailand for funding.
\end{acknowledgments}

%


\end{document}